\documentclass[prb,twocolumn,preprintnumbers,amsmath,amssymb,nofootinbib,floatfix]{revtex4}

\usepackage{graphicx, bm, tikz}
\usepackage[breaklinks]{hyperref}
\usepackage{mathrsfs}
\usepackage{braket}

\makeatletter
\def\graphicscale{\twocolumn@sw{0.3}{0.4}}
\def\graphicthreescale{\twocolumn@sw{0.3}{0.4}}

\begin{document}

\title{Quantum critical systems with dissipative boundaries}

\author{Francesco Tarantelli}
\affiliation{Dipartimento di Fisica dell'Universit\`a di Pisa,
        Largo Pontecorvo 3, I-56127 Pisa, Italy}

\author{Ettore Vicari} 
\affiliation{Dipartimento di Fisica dell'Universit\`a di Pisa
        and INFN, Largo Pontecorvo 3, I-56127 Pisa, Italy}

\date{\today}

\begin{abstract}
  We study the effects of dissipative boundaries in many-body systems
  at continuous quantum transitions, when the parameters of the
  Hamiltonian driving the unitary dynamics are close to their critical
  values. As paradigmatic models, we consider fermionic wires subject
  to dissipative interactions at the boundaries, associated with
  pumping or loss of particles. They are induced by couplings with a
  Markovian baths, so that the evolution of the system density matrix
  can be described by a Lindblad master equation. We study the quantum
  evolution arising from variations of the Hamiltonian and dissipation
  parameters, starting at $t=0$ from the ground state of the
  Hamiltonian at, or close to, the critical point. Two different dynamic
  regimes emerge: (i) an early-time regime for times $t\sim L$, where
  the competition between coherent and incoherent drivings develops a
  dynamic finite-size scaling, obtained by extending the scaling
  framework describing the coherent critical dynamics of the closed
  system, to allow for the boundary dissipation; (ii) a large-time
  regime for $t\sim L^3$ whose dynamic scaling describes the late
  quantum evolution leading to the $t\to\infty$ stationary states.
\end{abstract}

\maketitle


\section{Introduction}
\label{intro}

The out-of-equilibrium dynamics of quantum many-body systems has been
much investigated in the recent years.  The recent progress in quantum
technologies has also enabled experimental studies in the presence of
dissipation, either associated with unavoidable incoherent mechanisms,
or with suitably engineered system-bath couplings. Dissipative
mechanisms arising from the interaction with an
environment~\cite{HTK-12, MDPZ-12, CC-13, AKM-14} may lead to the
emergence of new collective phenomena, such as novel quantum phases and
phase transitions driven by
dissipation~\cite{Hartmann-16,NA-17,MBBC-18,LCF-19,SRN-19}, and
the emergence of dynamic scaling behaviors in the low-dissipative
regime of many-body systems at quantum
transitions~\cite{RV-21-rev,YMZ-14,YLC-16,NRV-19,RV-19,RV-20}.

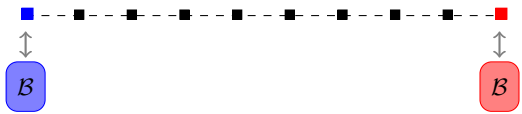
\begin{figure}[!b]
  \begin{tikzpicture}[scale=0.7]
    \draw [black,dashed] (1,0) -- (10,0);


    \foreach \i in {1}
{
        \filldraw [blue] (\i-0.07,-0.07) rectangle ++(6pt,6pt); 
        \draw[thick,<->,gray] (\i,-0.3)--(\i,-0.8);
        \draw [very thick, blue, rounded corners]
        (\i-0.35,-1.8) rectangle (\i+0.35,-0.9);
        \filldraw[blue!50!white, rounded corners]
        (\i-0.35,-1.8) rectangle (\i+0.35,-0.9);
        \node at (\i,-1.35) {$\mathcal{B}$};
}
    \foreach \i in {2,...,9}
{
        \filldraw [black] (\i-0.07,-0.07) rectangle ++(5pt,5pt); 
}
    \foreach \i in {10}
{
        \filldraw [red] (\i-0.07,-0.07) rectangle ++(6pt,6pt); 
        \draw[thick,<->,gray] (\i,-0.3)--(\i,-0.8);
        \draw [very thick, red, rounded corners]
        (\i-0.35,-1.8) rectangle (\i+0.35,-0.9);
        \filldraw[red!50!white, rounded corners]
        (\i-0.35,-1.8) rectangle (\i+0.35,-0.9);
        \node at (\i,-1.35) {$\mathcal{B}$};
}
  \end{tikzpicture}
  \caption{Sketch of a one-dimensional lattice system with boundary
    dissipation, arising from the interactions with two baths at the
    boundary sites, which may be of different nature.  In this paper
    we focus on the effects of boundary dissipative interactions when
    the quantum system is close to a bulk quantum transition, and
    therefore is characterized by quantum critical correlations.}
  \label{fig:sketchbou}
\end{figure}

In this paper we address the effects of dissipative boundaries in
many-body systems, such as the set up sketched in
Fig.~\ref{fig:sketchbou}, at continuous quantum transitions (CQTs),
when the parameters of the Hamiltonian driving the unitary dynamics
are close to their critical values.

Some issues related to the effects of dissipation on quantum systems
at CQTs have been already
investigated~\cite{YMZ-14,YLC-16,NRV-19,RV-19,RV-20}. We recall that
isolated many-body systems at CQTs develop dynamic scaling behaviors,
characterized by a diverging length scale $\xi$, and a vanishing gap
$\Delta$, as $\Delta\sim\xi^{-z}$ where $z$ is the universal dynamic
exponent~\cite{SGCS-97,Sachdev-book}.  The dissipative mechanisms
considered in Refs.~\onlinecite{YMZ-14,YLC-16,NRV-19,RV-19,RV-20} were
modeled by Lindblad master equations governing the time evolution of
the density matrix.  A dynamic scaling behavior emerges even in the
presence of dissipation, whose main features are controlled by the
universality class of the CQT.  However, such a dynamic scaling limit
requires a particular tuning of the dissipative interactions, whose
dissipative rate $u$ must scale as $u\sim \Delta \sim \xi^{-z}$.
These studies have been also extended to first-order quantum
transitions~\cite{DRV-20}, where a peculiar dynamic scaling emerges as
well, which appears more complex due to the strong sensitivity of
first-order transitions to the boundary conditions~\cite{CNPV-14}.

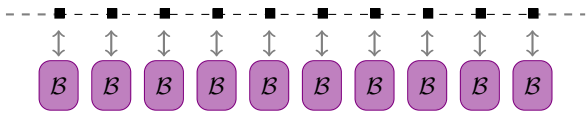
\begin{figure}[!b]
  \begin{tikzpicture}[scale=0.7]

    \draw [black,dashed] (1,0) --  (10,0);

    \draw [gray,thick,dashed] (0,0)--(1,0);
    \draw [gray,thick,dashed] (10,0)--(11,0);

    \foreach \i in {1,...,10}
{
        \filldraw [black] (\i-0.07,-0.07) rectangle ++(5pt,5pt); 
        \draw[thick,<->,gray] (\i,-0.3)--(\i,-0.8);
        \draw [very thick,violet, rounded corners]
        (\i-0.35,-1.8) rectangle (\i+0.35,-0.9);
        \filldraw[violet!50!white, rounded corners]
        (\i-0.35,-1.8) rectangle (\i+0.35,-0.9);
        \node at (\i,-1.35) {$\mathcal{B}$};
}
  \end{tikzpicture}
  \caption{Sketch of a one-dimensional lattice system in which the
    sites are homogeneously coupled to independent and equal baths
    ${\mathcal B}$, whose effect is to introduce local incoherent
    mechanisms. Therefore the dissipative mechanism is homogenous, and
    translation invariance is preserved (of course if translation
    invariance is satisfied by the Hamiltonian of the quantum system,
    and there are no boundaries, for example assuming periodic or
    antiperiodic boundary conditions). The effects of such homogeneous
    local dissipation at quantum transitions, modeled by corresponding
    Lindblad equations, have been analyzed in
    Refs.~\onlinecite{NRV-19,RV-19,RV-20}.}
  \label{fig:sketchhom}
\end{figure}

The above-mentioned works have focused on dissipative mechanisms
arising from homogenous couplings with external baths, involving the
bulk of the system, such as those sketched in
Fig.~\ref{fig:sketchhom}.  In this paper we consider a different
problem, focussing on critical systems subject to dissipative
interactions at the boundaries only, arising from environmental baths
that can only interact with the boundaries of the system, as sketched
in Fig.~\ref{fig:sketchbou}. We investigate the impact of boundary
dissipation to the quantum critical behavior of systems when it is
closed to its CQT, i.e. when the Hamiltonian parameters are tuned to a
quantum critical point.  We want to understand whether boundary
dissipations maintain the system within a critical regime, or they
make the system depart from criticality, whether their effects can be
casted within a dynamic scaling framework as in the case of homogenous
dissipative mechanisms.

We model the dissipative interaction with the environment by Lindblad
master equations for the density matrix of the system~\cite{BP-book,
  RH-book},
\begin{equation}
  {\partial\rho\over \partial t} = {\cal L}[\rho]:=
  -{i\over \hslash}\, [ \hat H,\rho]
  + {\mathbb D}[\rho]\,,
  \label{lindblaseq}
\end{equation}
where the generator ${\cal L}$ of the dynamics is called a Liouvillian
or a Lindbladian, the first term in the r.h.s. provides the coherent
Hamiltonian driving, while the second term accounts for the
dissipative coupling to the environment.  In the case of systems
weakly coupled to Markovian baths, the trace-preserving superoperator
${\mathbb D}[\rho]$ can be generally written as a sum of terms
associated with the various sources in contact with the
system~\cite{Lindblad-76, GKS-76}, i.e.
\begin{eqnarray}
  {\mathbb D}[\rho] &=& \sum_b w_b \, {\mathbb
    D}_b[\rho]\,, \label{drho}\\ {\mathbb D}_b[\rho] &=& \hat
  L_b^{\phantom\dagger} \rho \hat L_b^\dagger - \tfrac{1}{2} \big(
  \rho\, \hat L_b^\dagger \hat L_b^{\phantom\dagger} + \hat
  L_b^\dagger \hat L_b^{\phantom\dagger} \rho \big)\,, \nonumber
\end{eqnarray}
where $\hat L_b$ is the Lindblad jump operator describing the coupling
between the system and the bath labeled by $b$, and $w_b$ are
parameters controlling the strength or dissipative rate of the
corresponding dissipative interaction.  The form of the Lindblad
operators depends on the nature of the dissipation arising from the
interaction with the bath.  In quantum optical implementations, the
conditions leading to the Lindblad framework are typically
satisfied~\cite{SBD-16,DR-21}.

As a paradigmatic model, we consider the fermionic Kitaev
wire~\cite{Kitaev-01}, which undergoes a CQT belonging to the
two-dimensional Ising universality class. This choice allows us to
perform numerical computations for relatively large systems, thus
accurate checks of the dynamic scaling behaviors that we put forward.
We study the dynamic behavior close to the CQT, in the presence of
dissipation due to local incoherent particle pumping or loss at the
boundaries. We study the quantum evolution driven by the Lindbladian
master equation (\ref{lindblaseq}), arising from variations of the
Hamiltonian and dissipation parameters, starting at the initial time
$t=0$ from the ground state of the Hamiltonian close to the critical
point. We show that the quantum evolution of fermionic wires of size
$L$ is characterized by two distinct dynamic regimes. We observe an
early-time regime for $t\sim L$, where the competition between
coherent and incoherent drivings develop a dynamic finite-size scaling
(FSS), obtained by extending the dynamic FSS framework describing the
out-of-equilibrium critical dynamics of the closed
system~\cite{RV-21-rev}, to allow for the boundary dissipation.  Then,
at larger times, $t\sim L^3$, a different regime sets in, whose
dynamic scaling describes the late quantum evolution leading to the
$t\to\infty$ stationary states.  The above scenario is observed
keeping the boundary dissipative rate fixed, i.e. no tuning of the
dissipative rate turns out to be necessary, unlike the case of
homogenous dissipative mechanisms.

We finally mention that some issues concerning localized dissipative
interactions with the environment have been already discussed in the
literature, see
e.g. Refs.~\onlinecite{PP-08,Prosen-08,BGZ-14,Znidaric-15,VCG-18,FCKD-19,
  TFDM-19,
  BMA-19,KMS-19,SK-20,WSDK-20,FMKCD-20,Rossini-etal-20,AC-21}, and in
particular the case of quantum Ising chains with dissipative
interactions at the ends of the
chain~\cite{PP-08,Prosen-08,BGZ-14,Znidaric-15,VCG-18,SK-20}, mainly
focussing on the approach to the asymptotic large-time stationary
states.  In this paper we are interested in the whole dynamic behavior
from the early-time regime to the large-time regime leading to the
stationary states, addressing in particular the interplay with the
finite size of the system.

The paper is organized as follows.  In Sec.~\ref{Kitaevmo} we present
the paradigmatic fermionic Kitaev wire and the boundary dissipative
interactions with the environment that we consider in our study;
moreover we outline the dynamic protocol that we consider to address
the interplay between coherent Hamiltonian and incoherent boundary
dissipative drivings.  In Sec.~\ref{dynsca} we introduce the dynamic
FSS framework that allows us to describe the early-time regime of the
out-of-equilibrium quantum evolution associated with the protocol that
we consider. In Sec.~\ref{numres} we present our numerical results,
showing the emergence of different early-time and large-time
regimes, characterized by different time scales.  Finally in
Sec.~\ref{conclu} we summarize and draw our conclusions.  In
App.~\ref{appnum} we report some details of the numerical
computations.

\section{Fermionic wires subject to boundary dissipation}
\label{Kitaevmo}

\subsection{The Kitaev model}
\label{kitmod}

We consider fermionic quantum wires of $L$ sites, whose quantum
unitary dynamics is driven by the Kitaev Hamiltonian~\cite{Kitaev-01}
\begin{equation}
  \hat H_{\rm K} = - t \sum_{x=1}^{L-1} \big( \hat c_x^\dagger \hat
  c_{x+1}^{\phantom\dagger} +
  \delta \, \hat c_x^\dagger \hat c_{x+1}^\dagger+{\rm h.c.}
  \big) - \mu \sum_{x=1}^L \hat n_x \,,
  \label{kitaev2}
\end{equation}
where $\hat c_x$ is the fermionic annihilation operator on the $x^{\rm
  th}$ site of the chain, $\hat n_x\equiv \hat c_x^\dagger \hat
c_x^{\phantom\dagger}$ is the density operator, and $\delta>0$.  Note
that the Hamiltonian (\ref{kitaev2}) describe a chain with open
boundary conditions. In the following we set $\hslash =1$, $t=1$ as
the energy scale, and $\delta=1$.

The Hamiltonian~\eqref{kitaev2} can be mapped into a spin-1/2 XY
chain, by means of a Jordan-Wigner transformation, see, e.g.,
Ref.~\onlinecite{Sachdev-book}.  Fixing $\delta=1$, the corresponding
spin model is the quantum Ising chain
\begin{equation}
  \hat H_{\rm Is} = -\sum_{x=1}^{L-1} \hat \sigma^{(1)}_x \hat
  \sigma^{(1)}_{x+1} - g\, \sum_{x=1}^L \hat \sigma^{(3)}_x\,,
  \label{isham}
\end{equation}
$\hat \sigma^{(k)}_x$ being the Pauli matrices and $g=-\mu/2$.  In the
following we prefer to stick with the Kitaev quantum wire, because the
boundary dissipation that we consider is more naturally defined for
Fermi lattice gases, in terms of particle pumping and loss
mechanisms.

The Kitaev model undergoes a CQT at $\mu=\mu_c = -2$, independently of
$\delta$, between a disordered ($\mu<\mu_c$) and an ordered
($|\mu|<|\mu_c|$) quantum phase.  This transition belongs to the
two-dimensional Ising universality
class~\cite{Sachdev-book,RV-21-rev}, characterized by the length-scale
critical exponent $\nu=1$, related to the renormalization-group
dimension $y_\mu = 1/\nu=1$ of the Hamiltonian parameter $\mu$ (more
precisely of the difference $\bar{\mu} \equiv \mu-\mu_c$). This
implies that, approaching the critical point, the length scale $\xi$
of the critical quantum fluctuations diverges as $\xi \sim
|\bar{\mu}|^{-\nu}$ where $\bar{\mu}\equiv \mu-\mu_c$. The dynamic
exponent $z=1$ associated with the unitary quantum dynamics can be
obtained from the power law $\Delta\sim\xi^{-z}$ of the vanishing gap
with increasing $\xi$.

\subsection{Boundary dissipative mechanisms}
\label{dissmech}

We focus on the dynamic behavior of the Fermi lattice
gas~\eqref{kitaev2} close to its CQT, in the presence of boundary
dissipation mechanisms as described by the Lindblad
Eq.~(\ref{lindblaseq}).  We consider dissipative mechanisms associated
with the boundary sites of the chain, as sketched in
Fig.~\ref{fig:sketchbou}.  Within the Lindblad framework
(\ref{lindblaseq}), they are described by the dissipator
\begin{eqnarray}
  &&{\mathbb D}[\rho] =
  w_1 {\mathbb D}_{1}[\rho] + w_L {\mathbb D}_{L}[\rho]  \,,
\label{dissipator}\\
&& {\mathbb D}_{x}[\rho] = \hat L_{x}^{\phantom\dagger} \rho \hat L_{x}^\dagger -
  \tfrac{1}{2} \big( \rho\, \hat L_x^\dagger \hat L_{x}^{\phantom\dagger} + \hat
  L_x^\dagger \hat L_x^{\phantom\dagger} \rho \big)\,,\nonumber
\end{eqnarray}
where $\hat L_x$ denotes the Lindblad operator associated with the
system-bath coupling scheme. The strength of the boundary dissipative
mechanisms are controlled by the parameters $w_1$ and $w_L$, which are
related to the dissipative rates of the two processes.  The Lindblad
operators $\hat L_{1}$ and $L_L$ describe the coupling of the boundary
sites with the corresponding baths ${\mathcal B}$, see
Fig.~\ref{fig:sketchbou}. We consider different dissipation mechanisms
at the two ends of the chain, associated with particle losses and
pumping, respectively~\cite{HC-13, KMSFR-17, Nigro-19, NRV-19}:
\begin{equation}
  \hat L_{{1}} = \hat c_{1} \,, \qquad
  \hat L_{{L}} = \hat c_{L}^\dagger \,.
  \label{loppe}
\end{equation}

\subsection{The protocol}
\label{protocol}

To address the competition between coherent and boundary dissipative
drivings, we study the evolution after a quench of the Hamiltonian
parameters and dissipative interactions with the external
baths. Analogous protocols have been also considered to study the
effects of homogenous dissipative mechanisms preserving translation
invariance~\cite{NRV-19,RV-19,DRV-20}.  The protocol that we consider
is as follows.

\begin{itemize}

\item  
  The system starts from the ground state $|0,\bar{\mu}_i\rangle $ of
  $\hat H_K$ for a generic $\bar{\mu}_i\equiv \mu_i-\mu_c$,
  sufficiently small to stay within the critical regime.  Therefore
  the initial density matrix is given by
  \begin{equation}
    \rho_i\equiv \rho(t=0) = |
    0, \bar{\mu}_i\rangle \langle 0, \bar{\mu}_i|\,.
    \label{rhoi}
    \end{equation}

\item
  The out-of-equilibrium dynamics starts at $t=0$, arising from a
  sudden quench of the Hamiltonian parameter, from $\bar{\mu}_i$ to
  $\bar{\mu}\equiv \mu-\mu_c$, and the simultaneous turning on the
  dissipative interactions at the boundaries, as described by the
  boundary dissipator (\ref{dissipator}), with
  dissipation parameters $w_1$ and $w_L$.

\item
  The evolution of the quantum system, and in particular its density
  matrix, is described by the Lindblad master equation
  (\ref{lindblaseq}).

\end{itemize}

The time evolution is studied by monitoring a number of observables,
such as the particle density
\begin{equation}
D(x,t) = {\rm Tr}[\rho(t) \,\hat n_{x} ]\,,
\label{denisty}
\end{equation}
and the fermionic current
\begin{equation}
  J(x,t) =   {\rm Tr}[\rho(t) \,\hat j_{x} ]\,,
  \quad \hat{j}_x =
i\left( \hat c_{x+1}^{\dagger} \hat c_{x}^{\phantom\dagger}
- \hat c_{x}^\dagger \hat c_{x+1}^{\phantom\dagger}\right)\,,
    \label{current}
\end{equation}
where $\rho(t)$ is the time dependent density matrix driven by the
Lindblad equation with dissipative boundaries.
Moreover, we consider the fixed-time fermionic correlations
\begin{eqnarray}
  G_c(x,y,t) &=& {\rm Tr}[\rho(t) (
    \hat c_{x}^\dagger \hat
  c_{y}^{\phantom\dagger} + \hat c_{y}^\dagger \hat
  c_{x}^{\phantom\dagger})]\,,\label{ctf}\\ G_p(x,y,t) &=& {\rm
  Tr}[\rho(t) (\hat c_{x}^\dagger \hat c_{y}^\dagger + \hat
  c_{y}^{\phantom\dagger} \hat c_{x}^{\phantom\dagger})]\,.\label{ptf}
\end{eqnarray}
Note that translation
invariance does not hold due to the boundaries.

\section{Dynamic scaling behavior in the presence of dissipation}
\label{dynsca}

Before presenting the results for the problem addressed in the paper,
we briefly review some features of the dynamic scaling framework,
which we will exploit to characterize the dynamics of critical systems
with dissipative boundaries.  This approach was developed in
Refs.~\onlinecite{NRV-19,RV-19,RV-20,DRV-20}, extending the dynamic
scaling framework for isolated systems~\cite{SGCS-97,
  ZDZ-05,Polkovnikov-05,Dziarmaga-05,DOV-09,DGP-10,GZHF-10,
  CEGS-12,KCH-12,FDGZ-16,PRV-18,PRV-18-loc,v-18,RV-19-decoh,NRV-19-work}
(see also Ref.~\onlinecite{RV-21-rev} for a review of results on these
issues).

The initial $t=0$ conditions of the observables monitoring the dynamic
evolution is simply provided by their expectation values on the ground
state $| 0, \bar{\mu}_i \rangle$ of the many-body Hamiltonian $H_K$ at
the initial value $\bar{\mu}_i$, which can be obtained by using the
initial pure-state density matrix $\rho_i = | 0, \bar{\mu}_i\rangle
\langle 0, \bar{\mu}_i|$ in Eqs.~(\ref{ctf}) and(\ref{ptf}).  Close to
the quantum transition, i.e. when $|\bar{\mu}_i|\ll 1$, they develop
asymptotic FSS behaviors~\cite{SGCS-97,CPV-14,RV-21-rev}.  Their
scaling behavior is controlled by critical exponents
$\nu=y_\mu^{-1}=1$ and $z=1$ of the Ising universality class, and by
the renormalization-group dimension $y_c=1/2$ of the fermionic
operators $\hat c_x$ and $\hat
c^\dagger_x$~\cite{Sachdev-book,RV-21-rev}.  The initial ($t=0$)
ground-state fermionic correlations $G_{i,c/p}$ behave
as~\cite{CPV-14}
\begin{eqnarray}
&& G_{i}(x,y,\bar{\mu}_i,L) = L^{-2y_c} \left[ {\cal
      G}_{i}(X,Y,\kappa_i) +
    O(L^{-1})\right],\qquad\label{gcpsca}\\
  &&X = x/L\,,\qquad Y =
  y/L\,,\qquad \kappa_i = \bar{\mu}_i L^{y_\mu}\,,
  \label{fssscavar}
  \end{eqnarray}
in the large-$L$ limit keeping $X$, $Y$, and $\kappa_i$ fixed.  Note
that $G_{i,c/p}$, and the corresponding scaling functions ${\cal
  G}_{i,c/p}$, maintain the separate dependence on both space
variables $x$ and $y$, due to the presence of the boundaries.
Moreover the presence of boundaries gives also rise to the leading
$O(1/L)$ scaling corrections, which are absent in the case of systems
without boundaries~\cite{CPV-14}, such as systems with periodic or
antiperiodic boundary conditions.  The scaling corrections arising
from the leading irrelevant operator at the Ising fixed point are more
suppressed for the two-dimensional Ising universality
class~\cite{CCCPV-00,CHPV-02,CPV-14}, as $L^{-2}$.

The equilibrium (ground-state) FSS behavior of the particle density is
more complex, since the leading contribution comes from analytical
terms, while the scaling part is subleading. Indeed the equilibrium
ground-state particle density, corresponding to the initial condition
of the protocol considered behaves as~\cite{CPV-14}
\begin{eqnarray}
D(x,\bar{\mu}_i,L) \approx D_{\rm reg}(x,\bar{\mu}_i,L) +
L^{-y_n} {\cal D}(X,\kappa_i)\,,\quad
\label{dix}
\end{eqnarray}
where $y_n=d+z-y_\mu=1$ is the renormalization-group dimension of the
particle density operator.  The regular function $D_{\rm reg}$
provides the leading behavior, which arises from short-ranged
fluctuations, while the scaling part arising from the critical modes
is suppressed by a power $L^{-y_n}$. This does not make the particle
density particularly effective to highlight phenomena arising from
quantum long-range fluctuations. Fermionic correlations, such as those
in Eqs.~(\ref{ctf}) and (\ref{ptf}), are more suitable for this
purpose.  We also mention that the fermionic current $J(x,t)$,
cf. Eq.~(\ref{current}), vanishes at equilibrium, thus its initial
value is zero.

The equilibrium ground-state FSS can be extended to address
out-of-equilibrium coherent evolutions, for example arising from
instantaneous quenches of the Hamiltonian parameter from $\bar{\mu}_i$
to $\bar{\mu}$, starting from the ground state associated with
$\bar{\mu}_i$ at $t=0$. This requires the introduction of a further
scaling variable associated with time, given by $\theta = t L^{-z}
\sim t \Delta_L$, where $\Delta_L\sim L^{-z}$ is the gap (i.e., the
difference between the lowest energy levels) of the critical
Hamiltonian $H_K$ at $\bar{\mu}=0$.  For open boundary
conditions,~\cite{Pfeuty-70} $\Delta_L = {\pi/L} + O(L^{-2})$ at
$\bar{\mu}=0$.  The asymptotic dynamic FSS of the fermionic
correlations $G_{c/p}$, associated with a quench of the Hamiltonian
parameter from $\bar{\mu}_i$ to $\bar{\mu}$, can be written
as~\cite{PRV-18}
\begin{eqnarray}
&& G_{c/p}(x,y,\bar{\mu}_i,\bar{\mu},t,L) \approx L^{-2y_c} {\cal
    G}_{c/p}(X,Y,\kappa_i,\kappa,\theta)\,,\qquad
  \label{dyngcpsca} \\
    && \kappa = \bar{\mu}
 L^{y_\mu}\,,\qquad \theta= t L^{-z}\sim t \Delta_L\,,
 \label{dynfssscavar}
 \\
  &&y_c=1/2\,,\qquad y_\mu=1\,, \qquad z=1\,.\label{exponents}
\end{eqnarray}
Therefore, dynamic FSS in quenches from $\bar{\mu}_i$ to $\bar{\mu}$
is obtained in the large-$L$ limit keeping the scaling variables $X$,
$Y$, $\kappa_i$, $\kappa$, and $\theta$ fixed.

To monitor the out-equilibrium dynamics arising from the combination
of unitary Hamiltonian and incoherent dissipative drivings, it is
convenient to consider the rescaled correlation functions
\begin{eqnarray}
  \widetilde{G}_{c/p}(x,y,\bar{\mu}_i,\bar{\mu},\{w_b\},t,L)
  \equiv
  {  G_{c/p}(x,y,\bar{\mu}_i,\bar{\mu},\{w_b\},t,L)\over
     G_{i,c/p}(x,y,\bar{\mu}_i,L)}\,,\quad
 \label{rescagfunc} 
\end{eqnarray}
where $\{w_b\}$ indicates the dissipation parameters entering
Eq.~(\ref{drho}), and $G_{i,c/p}$ are the initial $t=0$ correlations.
Starting from $\widetilde{G}_{c/p}=1$ at $t=0$, they monitor the
variations of the fixed-time fermionic correlations from the initial
critical ground-state behavior.

In the case of homogenous couplings to the environment sources with
equal dissipator parameters
\begin{equation}
  w_b= u\quad {\rm for}\;\;b=1,...,L\,,
  \label{wbu}
  \end{equation}
associated with identical local baths, such as those sketched in
Fig.~\ref{fig:sketchhom}, one observes the emergence of a dynamic
scaling regime as well~\cite{NRV-19,RV-19,RV-21-rev}, involving a
further scaling variables associated with the dissipation parameters
$w$.  The analysis of
Refs.~\onlinecite{YMZ-14,YLC-16,NRV-19,RV-19,RV-20} shows that
dissipation represents a relevant perturbation at CQTs, leading out of
criticality similarly to the temperature. Thus an appropriate tuning
is required to stay within the critical regime.  This is achieve by
considering the scaling variable
\begin{equation}
\gamma = u L^z\sim u/\Delta_L\,,\qquad z=1\,.
\label{gammadef}
\end{equation}
Then, the dynamic FSS behavior of the
fermionic correlations reads~\cite{NRV-19,RV-19}
\begin{eqnarray}
 \widetilde{G}_{c/p}(x,y,\bar{\mu}_i,\bar{\mu},u,t,L) \approx
  \widetilde{\cal G}_{c/p}(X,Y,\kappa_i,\kappa,\theta,\gamma)\,.\quad
  \label{gcpscadiss}
\end{eqnarray}
Therefore the dynamic FSS behaviors in the presence of homogenous
dissipation is asymptotically obtained in the large-$L$ limit keeping
also the scaling variable $\gamma$ fixed. In particular, this implies
that the Hamiltonian parameters $\bar\mu_i$ and $\bar\mu$ must remain
close to the critical value $\bar{\mu}_i=\bar{\mu}=0$, and the
dissipation parameter must be tuned to low values, i.e., $u\sim
L^{-z}$, to remain within the critical regime during the time
evolution. Dynamic scaling laws in the thermodynamic limit can be
obtained from the above FSS laws~\cite{RV-19,RV-21-rev}, by taking the
limit $L/\lambda\to\infty$ where $\lambda=|\bar{\mu}|^{-\nu}$
represents a length scale.  The above dynamic scaling behaviors have
been accurately checked within the Kitaev model with antiperiodic
boundary conditions and homogenous local couplings to baths associated
with pumping, decay and dephasing~\cite{NRV-19,RV-19}.  The asymptotic
dynamic scaling behavior is generally approached with $O(1/L)$ or
$O(\lambda^{-1})$ corrections~\cite{RV-19}.  The above
studies~\cite{NRV-19,RV-19} considered dissipative systems without
boundaries, assuming antiperiodic boundary conditions, for which
translation invariance is preserved even for finite systems. We have
verified that the dynamic FSS (\ref{gcpscadiss}) is also
asymptotically observed when considering Kitaev wires with open
boundary conditions (some results are shown later), thus in the
presence of boundaries.

In this paper we consider the case of dissipative interactions with
external sources limited to the boundaries of the system.  We again
exploit an analogous dynamic FSS framework to discuss the relevance of
boundary dissipation at CQTs. We recall that, in the case of closed
systems, the boundary conditions do not change the universal power
laws of the dynamic FSS, but only the scaling functions depend on
them. Here we want to understand what happens in the presence of
dissipative boundaries, in particular under which condition they
maintain the system within the critical regime, and the main features
of the quantum evolution in their presence.

For systems with dissipative boundaries we put forward dynamic FSS
behaviors similar to that holding for homogenous dissipative
mechanisms, see Eq.~(\ref{gcpscadiss}).  For simplicity, we consider
the following cases: 
\begin{eqnarray}
&{\rm (i)}\quad &w_1=w_L=w\,, \label{w1lw}\\
&{\rm (ii)} \quad &w_1=w\,, \quad w_L=0\,,  \nonumber
\end{eqnarray}
describing respectively pumping/loss dissipation at the boundaries
with equal strength $w$, and loss dissipation at one boundary only.
Note that in both cases we use the same symbol $w$ for the dissipation
rate.  Our working hypothesis for both cases in Eqs.~(\ref{w1lw}) is
that the early-time $t\sim L$ dynamics of fermionic correlations
asymptotically develops the dynamic FSS
\begin{eqnarray}
 \widetilde{G}_{c/p}(x,y,\bar{\mu}_i,\bar{\mu},w,t,L) \approx
 \widetilde{\cal G}_{c/p}(X,Y,\kappa_i,\kappa,\theta,wL^\zeta)\,,\quad
\label{gcpscadiss3}
\end{eqnarray}
where $\zeta$ is further exponent characterizing the relevance of the
dissipative boundaries at the CQT of the closed system.  In the next
section, we will provide numerical evidence of such dynamic FSS,
supporting also the absence of rescaling of the dissipative parameter,
i.e. $\zeta=0$. The more general $w_1\neq w_L$ case can be
straightforwardly addressed by considering separate dependences on
$w_1L^{\zeta_1}$ and $w_LL^{\zeta_L}$ in the scaling functions
$\widetilde{\cal G}_{c/p}$.

\section{Numerical results}
\label{numres}

We now present our numerical results for the fermionic Kitaev chain
with dissipative boundaries. We mostly focus on the the case (i) of
Eqs.~(\ref{w1lw}), with decay and pumping dissipative interactions at
the ends $x=1$ and $x=L$ respectively. We also report some results for
the case (ii) of Eqs.~(\ref{w1lw}) with only the decay-type
dissipation at one end.  Details of the computations are reported in
App.~\ref{appnum}.

\subsection{Time evolution and asymptotic stationary states}
\label{statstates}

\begin{figure}[!t]
  \includegraphics[width=0.95\columnwidth]{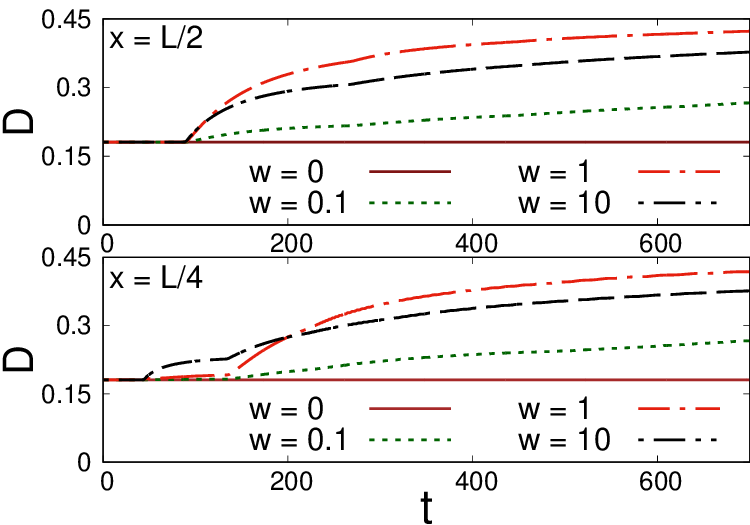}
  \caption{The time evolution of the particle density $D(x,t)$ arising
    from the protocol outlined in Sec.~\ref{protocol}, in the presence
    of pumping/decay boundary dissipative mechanisms, for
    $\bar{\mu}_i=\bar{\mu}=0$, $L=360$, $x=L/4$ (bottom) and $x=L/2$
    (top), and various values of $w$, versus the time $t$.  Note that
    it takes some time, $t=t^*>0$ to depart from the initial value,
    corresponding to the time needed for a signal of speed $v_m=2$ to
    travel from the closest end to the site $x$, see text.  }
  \label{Dt}
\end{figure}

\begin{figure}[!t]
  \includegraphics[width=0.95\columnwidth]{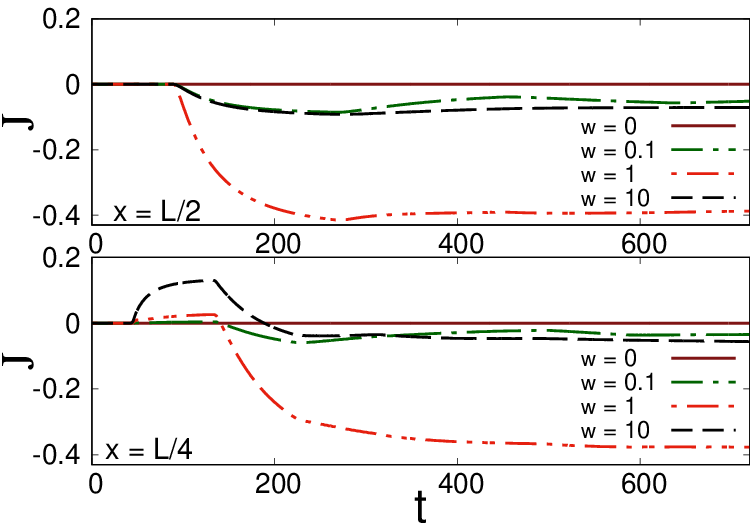}
  \caption{The time evolution of the fermionic current $J(x,t)$, for
    $\bar{\mu}_i=\bar{\mu}=0$, $L=360$, $x=L/4$ (bottom) and $x=L/2$
    (top), and various values of $w$, versus the time $t$.  In the
    presence of pumping/decay boundary dissipative mechanisms, the
    fermionic current becomes nonzero.  Of course, the current goes
    from the end where the bath is pumping particles to the other one
    where they get lost.}
  \label{Jt}
\end{figure}

\begin{figure}[!t]
  \includegraphics[width=0.95\columnwidth]{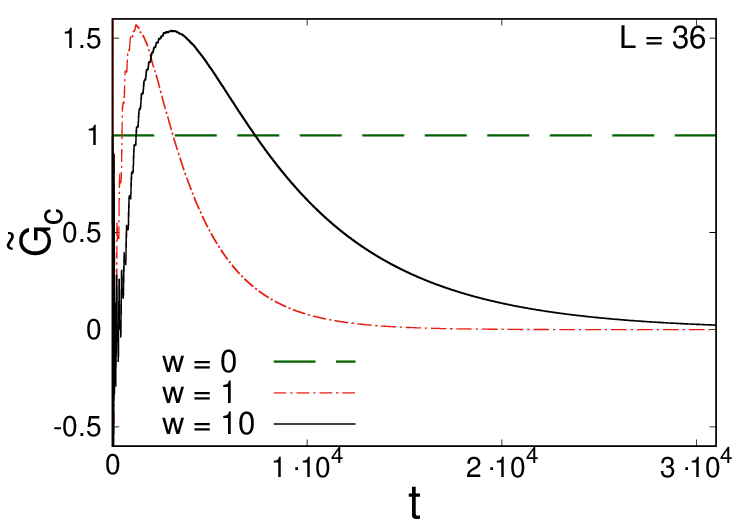}
    \includegraphics[width=0.95\columnwidth]{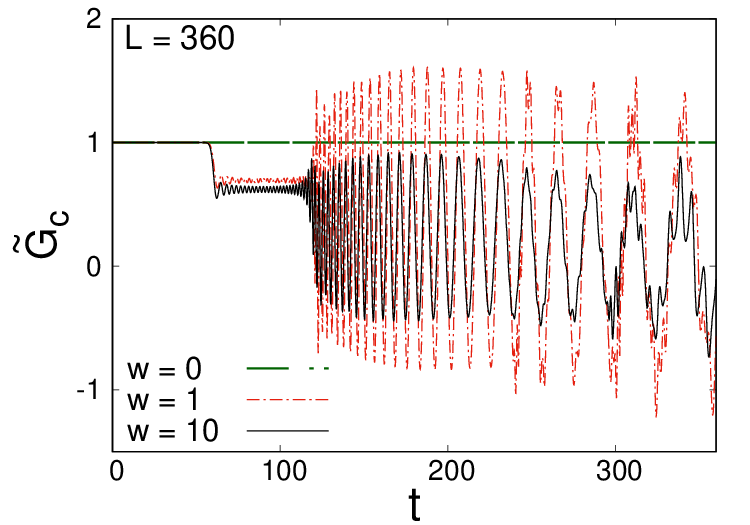}
    \caption{The time volution of the ratio $\widetilde{G}_c$,
      cf. Eq.~(\ref{rescagfunc}), related to the fixed-time fermionic
      correlation $G_{c}$, in the presence of pumping/decay boundary
      dissipative mechanisms, for $\bar{\mu}_i=\bar{\mu}=0$, $x=L/3$
      and $y=2L/3$ (symmetric with respect to the center of the
      lattice), and various values of $w$, versus the time $t$.  The
      top figure shows data for $L=36$ up to large times approaching
      the stationary state. The bottom figure shows data for $L=360$
      and relatively small time.  In the presence of dissipation
      $w>0$, we note a significant change of behavior, being
      characterized by ample oscillations, whose frequency appear
      approximately independent of $w$, while the oscillation
      amplitudes are apparently nonmonotonic, being larger for $w=1$
      than $w=10$.}
    \label{Gcpt}
\end{figure}

\begin{figure}[!t]
  \includegraphics[width=0.95\columnwidth]{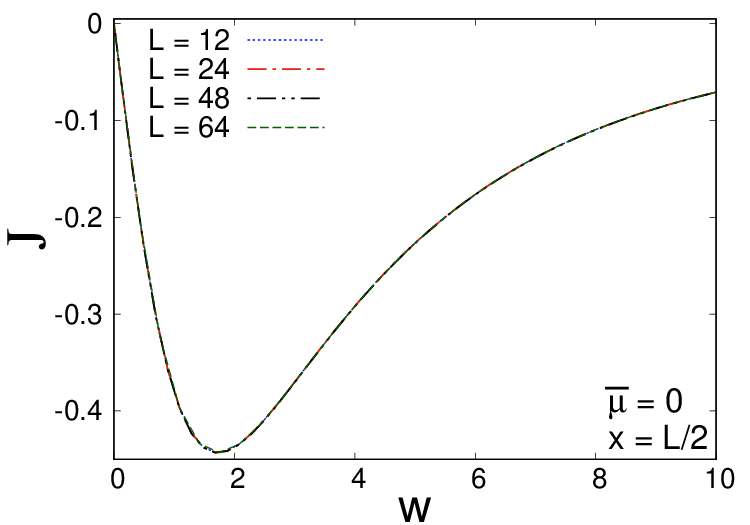}
  \caption{Asymptotic large-$t$ values of $J(x,t)$ for $\bar{\mu}=0$
    and $x=L/2$ for various sizes $L$. They turn out to be hardly
    distinguishable.  The same results are obtained for any site not
    involving the ends of the chain. These results provide the
    dependence on $w$ of the asymptotic large-time limit
    $f(\bar{\mu},w)$ of the fermionic current $J(x,t)$,
    cf. Eq.~(\ref{jasy}), for $\bar{\mu}=0$. Note that $f(0,w)$ is
    nonmonotonic, indeed its absolute value has a maximum at a finite
    value $w\approx 1.7265$.}
  \label{Jwlt}
\end{figure}

In Figs.~\ref{Dt}, \ref{Jt} and \ref{Gcpt}, we show some results
for the evolution of the particle density, the fermionic current, and
the correlation $G_c$ respectively, for protocols with pumping and
decay dissipative boundaries, when starting from a critical ground
state at $\bar{\mu}_i=0$. The quantum evolution leads to stationary
states depending on $\bar{\mu}$ and $w$, while it is independent of
the initial condition, thus independent of $\bar{\mu}_i$. The
asymptotic stationary state corresponds to the eigenstate of the
Lindbladian generator ${\cal L}$ with zero eigenvalue, i.e. it is
solution of ${\cal L}(\rho)=0$. The observables defined in
Sec.~\ref{protocol} asymptotically turn out to behave as
\begin{eqnarray}
  && {\rm lim}_{t\to\infty} \; D(x,\bar{\mu}_i, \bar{\mu},w, t,L) = {1/2}\,,
  \label{dasy}\\
  && {\rm lim}_{t\to\infty} \; J(x,\bar{\mu}_i,
  \bar{\mu},w, t,L) = f(\bar{\mu},w)\,,
  \label{jasy}\\
  &&{\rm lim}_{t\to\infty} \; G_{c/p}(x,y,\bar{\mu}_i, \bar{\mu},w, t,L) = 0\,,
  \label{gcpasy}
  \end{eqnarray}
for any sites except those at the ends of the chain that are in
contact with the baths (the fermionic density and current take
different values at the boundaries).  Only the asymptotic large-$t$
fermionic current given by the function $f(\bar{\mu},w)$ shows a
behavior dependent on $\bar{\mu}$ and $w$, see Fig.~\ref{Jwlt}, being
nonzero and constant for any site $x$ excluded those involving the
ends of the chain. The above results can be derived by solving the
corresponding dynamic equations, see App.~\ref{appnum}, in the
stationary limit when the time derivatives in the l.h.s. vanish.

The asymptotic stationary states do not appear particularly
interesting.  However, we are mostly interested in the quantum
evolution before approaching the asymptotic stationary states. As we
shall see, this turns out to be quite complex, developing two
different dynamic regimes: an early-time regime for $t\sim L$, and a
large-time regime for $t\sim L^3$ that describe the approach to the
above stationary states.

We note that in protocols without quenching of the Hamiltonian
parameters, thus limiting itself to switch the boundary dissipative
interactions on, the observables far from the ends remain unchanged up
to a certain time $t=t^*>0$, see Figs.~\ref{Dt}, \ref{Jt} and
\ref{Gcpt} (all obtained without quenching the Hamiltonian parameter
$\bar{\mu}$).  This fact can be related to the propagation of the
quasi-particle modes within the bulk of the system~\cite{LR-72,CC-05}.
In the equivalent quantum Ising chain, cf. Eq.~(\ref{isham}), their
maximum speed is given by $v_m = 2 \, {\rm Min}
[g,1]$,~\cite{CEF-12-1} therefore $v_m=2$ at the critical point.  For
example Fig.~\ref{Dt} shows that the particle density at $x=L/2$ and
$x=L/4$ starts departing from its initial value at $t^*\approx L/4$
and $t^*=L/8$ respectively, which is the time needed for a signal of
speed $v_m=2$ to arrive at the site $x$, starting from the closest
dissipative end at $t=0$. Analogous initial behaviors are observed for
the other observables considered.

We also note that the time scale of the signal propagation,
i.e. $t\sim L$, is compatible with the time scaling variable
$\theta=t/L$ introduced in Sec.~\ref{dynsca}. Therefore, we expect
that phenomena related to propagation are essentially encoded in the
asymptotic dynamic scaling functions entering Eq.~(\ref{gcpscadiss}).
We also mention that the finite-speed propagation of quasi-particle
modes gives also rise to peculiar revival phenomena in closed
finite-size systems~\cite{RV-21-rev,HHH-12, KLM-14, Cardy-14, JJ-17,
  MAC-20,RV-21}.

\subsection{The early-time dynamic finite-size scaling}
\label{earlytimeFSS}

\begin{figure}[!t]
    \includegraphics[width=0.95\columnwidth]{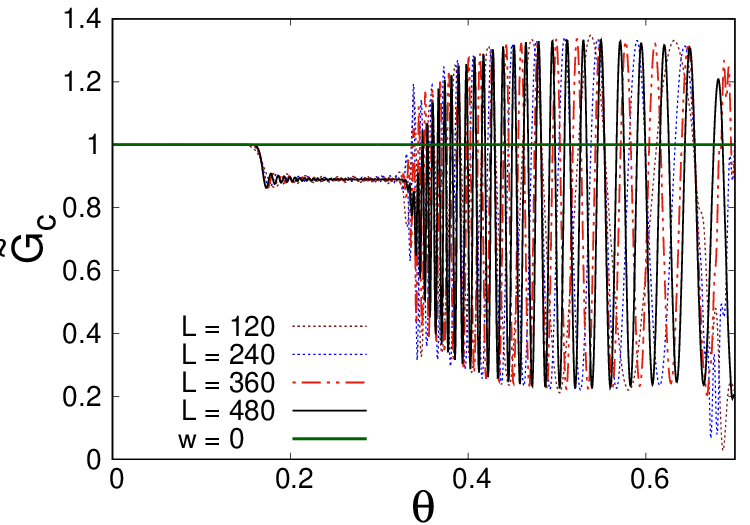}
    \caption{ Dynamic scaling of the ratio
      $\widetilde{G}_{c}(x,y,\bar{\mu}_i,\bar{\mu},w,t,L)$ associated
      with the fermionic correlation $G_c$, as defined in
      Eq.~(\ref{rescagfunc}), for fixed $X=x/L=1/3$, $Y=y/L=2/3$ (so
      that $Y-X=1/3$), $\kappa_i=\kappa=0$, $w=1/4$, versus $\theta =
      t/L$, for various size $L$ up to $L=480$.  These results support
      the dynamic FSS put forward in Eq.~(\ref{gcpscadiss3}), indeed
      they appear to converge to an asymptotic scaling function of
      $\theta$, characterized by sizeable oscillations whose amplitude
      and frequency appear to scale as well. The approach to the
      asymptotic behavior is globally compatible with the existence of
      $O(1/L)$ corrections. The case for $w=0$, i.e. no dissipation,
      is trivially represented by a constant line,
      $\widetilde{G}_c=1$.}
     \label{Gcw}
\end{figure}

\begin{figure}[!t]
  \includegraphics[width=0.95\columnwidth]{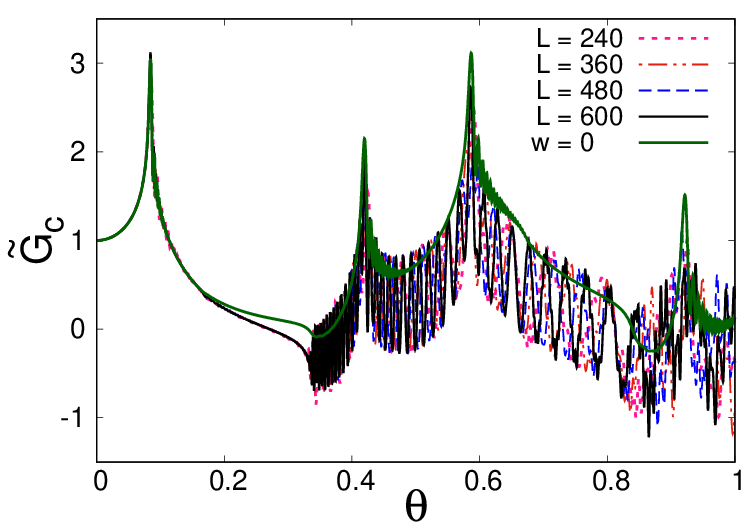}
  \includegraphics[width=0.95\columnwidth]{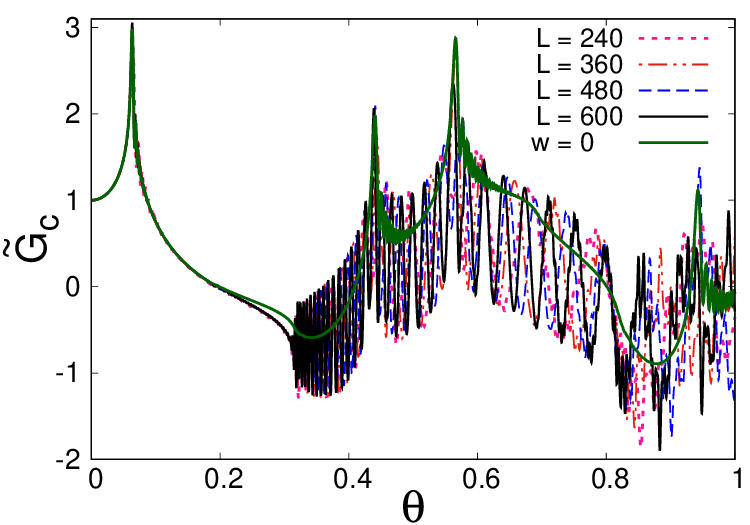}
 \caption{ Dynamic scaling of
   $\widetilde{G}_{c}(x,y,\bar{\mu}_i,\bar{\mu},w,t,L)$ for fixed
   $X=x/L=1/3$, $Y=y/L=2/3$ (so that $Y-X=1/3$, top figure) and
   $X=x/L=3/8$, $Y=y/L=5/8$ (so that $Y-X=1/4$, bottom figure),
   $\kappa_i=0$, $\kappa=3$, $w=1/4$, versus $\theta = t/L$, for
   various size $L$ up to $L=600$.  These results support the dynamic
   FSS put forward in Eq.~(\ref{gcpscadiss3}). The comparison with the
   case without dissipation, i.e. the $w=0$ curve for $L=480$, show
   similarities, but very distinct oscillatory behaviors for $w=1/4$.
 }
  \label{Gcwq}
\end{figure}

\begin{figure}[!t]
  \includegraphics[width=0.95\columnwidth]{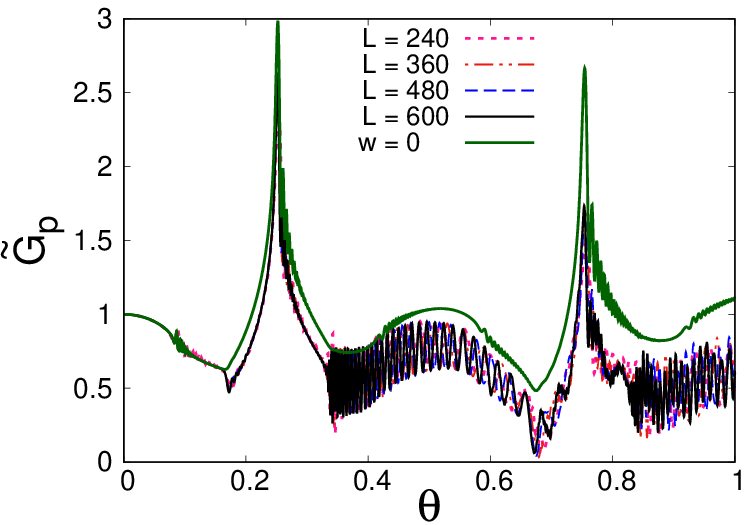}
  \includegraphics[width=0.95\columnwidth]{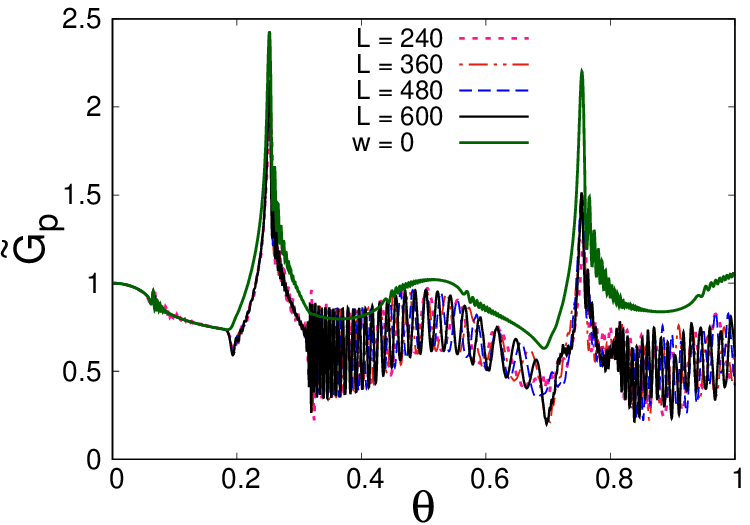}
 \caption{ Dynamic scaling of the ratio
   $\widetilde{G}_{p}(x,y,\bar{\mu}_i,\bar{\mu},w,t,L)$ associated
   with the fermionic correlation $G_p$, cf.  Eq.~(\ref{rescagfunc}),
   for fixed $X=x/L=1/3$, $Y=y/L=2/3$ (so that $Y-X=1/3$, top figure)
   and $X=x/L=3/8$, $Y=y/L=5/8$ (so that $Y-X=1/4$, bottom figure),
   $\kappa_i=0$, $\kappa=3$, $w=1/4$, versus $\theta = t/L$, for
   various size $L$ up to $L=600$.  These results support the dynamic
   FSS put forward in Eq.~(\ref{gcpscadiss3}). For comparison we also
   report the curve for $w=0$ for $L=480$.}
  \label{Gpwq}
\end{figure}

\begin{figure}[!t]
 \includegraphics[width=0.95\columnwidth]{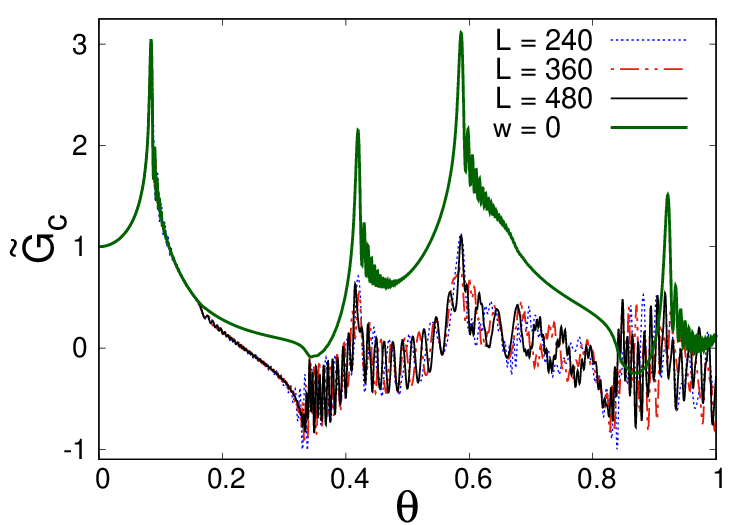}
  \caption{Dynamic scaling of
    $\widetilde{G}_{c}(x,y,\bar{\mu}_i,\bar{\mu},w,t,L)$ for a single
    decay dissipative boundary, for $X=x/L=1/3$ and $Y=y/L=2/3$, at
    fixed $\kappa_i=0$, $\kappa=3$, and $w=10$, versus $\theta = t/L$,
    for various size $L$ up to $L=480$. Again the dynamic FSS
    (\ref{gcpscadiss3}) emerges, thus supporting the value $\zeta=0$
    for the exponent entering the scaling variable associated with
    $w$.  We again compare with the case in the absence of
    dissipation, i.e. for $w=0$ (the reported curve is that for
    $L=480$). }
      \label{Gcsbw}
\end{figure}

\begin{figure}[!t]
  \includegraphics[width=0.95\columnwidth]{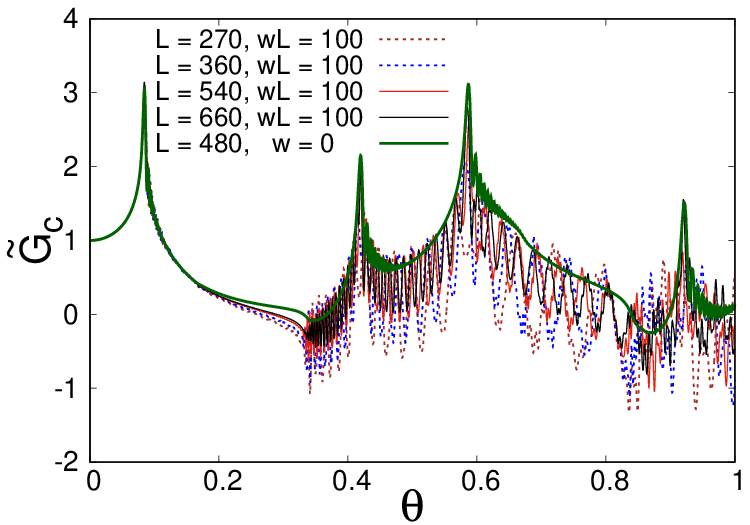}
 \caption{Check of the behavior of the fermionic correlation
   $\widetilde{G}_{c}(x,y,\bar{\mu}_i,\bar{\mu},w,t,L)$ when
   decreasing the dissipation parameter as $w \sim L^{-1}$, like the
   dynamic scaling for homogenous dissipations. In particolar, we
   report results for fixed $X=x/L=1/3$, $Y=y/L=2/3$ (so that
   $Y-X=1/3$), $\kappa_i=0$, $\kappa=3$, $w L=100$, versus $\theta =
   t/L$, for various size $L$ up to $L=660$. The curves appear to
   approach that in the absence of dissipation, with oscillations
   whose amplitude is decreasing as $1/L$ approximately.
  }
  \label{GcwL}
\end{figure}

\begin{figure}[!t]
      \includegraphics[width=0.95\columnwidth]{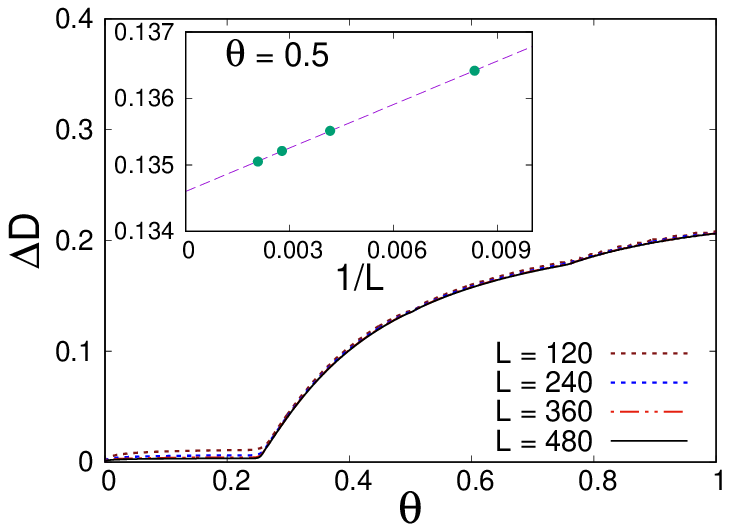}
    \caption{Dynamic scaling of the subtracted particle density
      $\Delta D(x,\bar{\mu}_i,\bar{\mu},w,t,L)$ defined in
      Eq.~(\ref{deltade}), at fixed $\kappa_i=0$, $\kappa=3$, and
      $w=1$ for various $L$. 
    }
  \label{rhow}
\end{figure}

We now investigate the early-time regime of the quantum evolution
arising from the protocol described in Sec.~\ref{protocol}. This is
the regime where $t\sim L$, therefore the appropriate time scaling
variable is $\theta=t/L$, like the dynamic FSS developed by closed
fermionic Kitaev wires at CQTs, cf. Eq.~(\ref{dyngcpsca}).

To determine the correct scaling associated with the boundary
dissipation parameter, and in particular the exponent $\zeta$ in
Eq.~(\ref{gcpscadiss3}), we compute the time evolution of the
fixed-time fermionic correlations $G_c$ and $G_p$ for various values
of $w$. Figs.~\ref{Gcw}, \ref{Gcwq} and \ref{Gpwq} show results at
fixed $w=1/4$.  We show results without quenching the Hamiltonian
parameter $\mu$, in Fig.~\ref{Gcw}, and quenching it around the
critical point, in Figs.~\ref{Gcwq} and \ref{Gpwq}.  Analogous results
are observed for other values of $w$.  The curves appear to approach
an asymptotic scaling behavior matching the FSS ansatz
(\ref{gcpscadiss3}), thus supporting the value $\zeta=0$ for the
exponent entering the scaling variable associated with $w$.  The
asymptotic scaling functions appear clearly distinct from those in the
absence of dissipation, i.e. for $w=0$.  Their comparison shows some
similarities, for example the existence of spikes, but very distinct
oscillatory behaviors for $w=1/4$, which persist in the dynamic FSS
limit. The convergence to the large-$L$ asymptotic behavior is
generally consistent with $O(1/L)$ corrections.  However, the
convergence is expected to be nonuniform, i.e. the amplitudes of the
corrections are expected to increase, making it slower and slower with
increasing $\theta$, as also shown by the data.  Analogous results are
also obtained in the case of a single decay dissipative boundary, see
Fig.~\ref{Gcsbw} for results with $w=10$.

As a check of the apparent dynamic scaling with $\zeta=0$, in
Fig.~\ref{GcwL}) we show plots obtained by keeping the product $w\,L$
fixed when increasing the size $L$, thus by decreasing the dissipation
parameter as $w\sim 1/L$. This is the scaling analogous to the case of
homogenous dissipators, cf. Eqs.~(\ref{gammadef}) and
(\ref{gcpscadiss}).  In this case, the curves appear to approach the
scaling function of the close system for $w=0$, and the oscillations
gets suppressed as $1/L$.

Therefore, we conclude that the dynamic FSS developed by the fermionic
correlations $G_{c/p}$ within the early-time regime is compatible with
a vanishing exponent $\zeta$ in Eq.~(\ref{gcpscadiss3}), and exclude
the value $\zeta=1$. Of course our numerical analysis cannot really
distinguish $\zeta=0$ from a small value, say $|\zeta|\lesssim 0.2$. A
more conclusive evidence for $\zeta=0$ would require exact
computations in the dynamic FSS limit, or numerical results for much
larger chains.  A simple (likely naive) interpretation of the evidence
in favor of $\zeta=0$ may be related to the fact that the dynamic FSS
for homogenous bulk dissipation requires $w\sim L^{-z}$, but it
involves a number $L$ of dissipators as in Fig.~\ref{fig:sketchhom}.
On the other hand, the boundary dissipation arises from a number of
dissipators smaller by a $O(1/L)$ factor.  Therefore, one might
interpret the vanishing of $\zeta$ as the result of the simple
relation $\zeta=z-1=0$. We believe that this point deserve further
investigation, for example by checking it in other models, with CQTs
characterized by different critical exponents.

We also show some results for the particle density, in particular for
the difference
\begin{equation}
  \Delta D(x,t) = D(x,t) - D(x,t=0)\,,
  \label{deltade}
  \end{equation}
see Fig.~\ref{rhow}.  They show an apparent scaling behavior when
plotted versus $\theta$, demonstrating that the early-time scale of
the variations of the particle density is $t\sim L$ as well. However,
this scaling behavior does not come from the original critical modes,
since their contributions are suppressed as $O(L^{-y_n})$, thus as
$O(L^{-1})$, see Eq.~(\ref{dix}). Analogous results are obtained for
the fermionic current.

We finally note that the asymptotic dynamic scaling behaviors of the
correlations, reported in Figs.~\ref{Gcw}-\ref{GcwL}, are
characterized by the presence of cusps, thus indicating a nonanalytic
time dependence in the rescaled time variable $\theta$. This features
are reminiscent of the behavior at the so-called dynamical phase
transitions~\cite{HPK-13,Heyl-18}. Likely, they deserve further
investigation.

\subsection{Large-time regime approaching
  stationary states}
\label{apprstatstates}

The approach to the asymptotic stationary states are generally
controlled by the Liouvillian gap $\Delta_{\cal L}$ associated with
the generator ${\cal L}$,
cf.
Eq.~(\ref{lindblaseq}).~\cite{BP-book,RH-book,Znidaric-15,MBBC-18,SK-20}.
The asymptotic stationary state is provided by the eigenstate of
${\cal L}$ with vanishing eigenvalue, $\Lambda_0=0$, while all other
eigenstates have eigenvalues $\Lambda_i$ with negative real part,
i.e. ${\rm Re}\,\Lambda_i<0$ for any $i>0$. The approach to the
stationary state is controlled by Liouvillian gap $\Delta_{\cal L}$
which is given by the eigenvalue with the largest nonzero real part,
i.e.
\begin{equation}
  \Delta_{\cal L} = - {\rm Max}_{i>0} \, {\rm
    Re}\,(\Lambda_i)\,.
    \label{deltadeb}
\end{equation}

\begin{figure}[!t]
  \includegraphics[width=0.95\columnwidth]{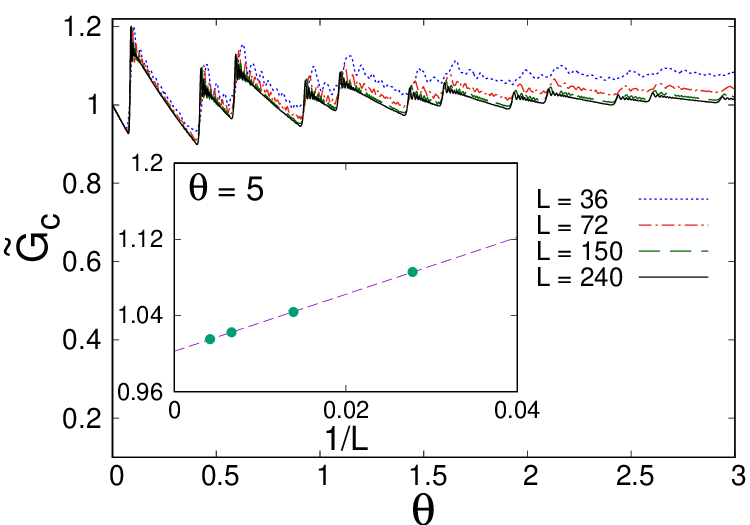}
  \caption{Evolution of
    $\widetilde{G}_{c}(x,y,\bar{\mu}_i,\bar{\mu},u,t,L)$ for the
    homogenous decay dissipation mechanisms, such as those sketched in
    Fig.~\ref{fig:sketchhom}, and for $X=x/L=1/3$ and $Y=y/L=2/3$, for
    $\bar{\mu}_i=\bar{\mu}=0$, and $\gamma=u L=1$ fixed, versus
    $\theta = t/L$, for various size $L$ up to $L=240$. The inset
    shows the $1/L$ convergence at $\theta=5$.  }
      \label{Gchomasy}
\end{figure}

As shown in Ref.~\onlinecite{RV-19}, in the case of homogenous
dissipative schemes (like that in Fig.~\ref{fig:sketchhom}), the
conjectured dynamic scaling, such as that in Eq.~(\ref{gcpscadiss}),
describes also the approach to the asymptotic stationary
states. Indeed, for homogenous local dissipative mechanisms such as
pumping or decay, $\Delta_{\cal L}$ scales as $\Delta_{\cal L} \sim
1/L$ when keeping $\gamma=uL$ fixed, analogously to the critical gap
$\Delta_L\sim 1/L$ at the CQT of the Kitaev wire.  Therefore, the
dynamic scaling can follow the whole dynamic process from $t=0$ to the
asymptotic stationary states. This is supported by the data in
Fig.~\ref{Gchomasy}, which show that the dynamic FSS ansatz
(\ref{gcpscadiss}) describes the whole quantum dynamics from $t=0$ to
the corresponding asymptotic stationary states. Note that
Fig.~\ref{Gchomasy} reports results for fermionic wires with open
boundary conditions, thus extending the evidence of dynamic FSS
already reported in Refs.~\onlinecite{NRV-19,RV-19} for systems with
antiperiodic boundary conditions.

\begin{figure}[!t]
  \includegraphics[width=0.95\columnwidth]{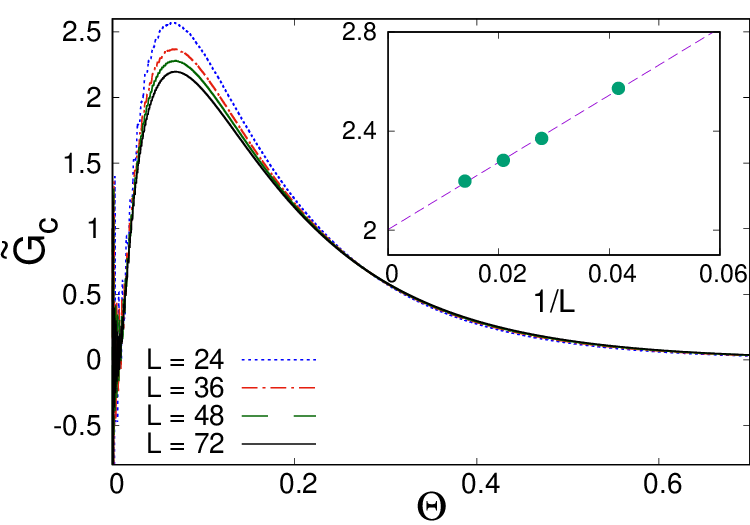}
    \includegraphics[width=0.95\columnwidth]{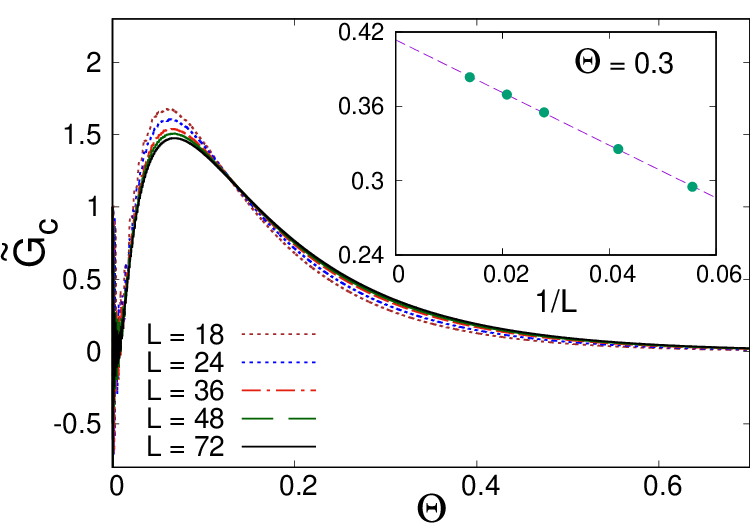}
  \caption{Evolution of
    $\widetilde{G}_{c}(x,y,\bar{\mu}_i,\bar{\mu},w,t,L)$ for the case
    with pumping/decay boundary dissipations, and for $X=x/L=1/3$ and
    $Y=y/L=2/3$, for $\kappa_i=\kappa=0$ (bottom) and
    $\kappa_i=\kappa=3$ (top), and $w=10$, versus $\Theta = t/L^3$,
    for various size $L$ up to $L=72$. The inset in the top figure
    shows the large-$L$ convergence at the maximum of the curve, while
    that of the bottom figure shows the convergence at $\Theta=0.3$.}
      \label{Gcwasy}
\end{figure}

\begin{figure}[!t]
  \includegraphics[width=0.95\columnwidth]{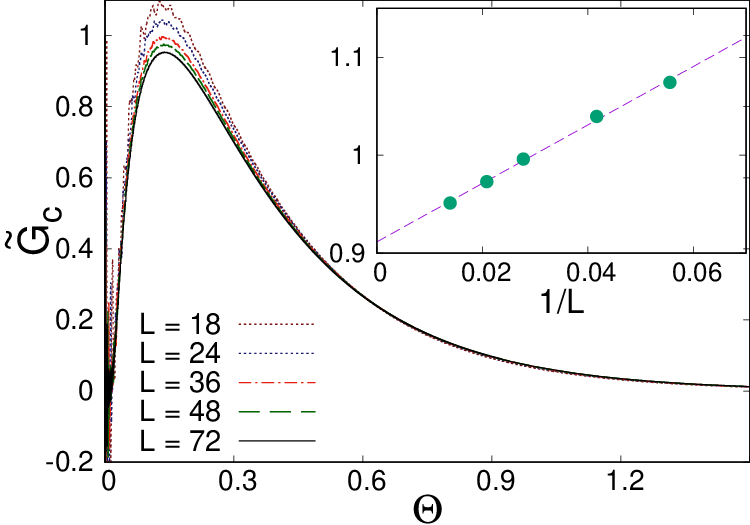}
  \caption{Evolution of
    $\widetilde{G}_{c}(x,y,\bar{\mu}_i,\bar{\mu},w,t,L)$ with a single
    decay dissipative boundary, for $X=x/L=1/3$, $Y=y/L=2/3$,
    $\bar{\mu}_i=\bar{\mu}=0$, and $w=10$, versus $\Theta = t/L^3$,
    for various size $L$ up to $L=72$.  The inset shows the large-$L$
    convergence at the maximum of the curves.}
      \label{Gcwasy1b}
\end{figure}

In the case of dissipative boundaries we observe another scenario,
for which the approach to the stationary behavior requires a different
scaling regime, characterized by much larger times, scaling as $t\sim
L^3$, instead of $t\sim L$. This is in agreement with analytical and
numerical results for quantum spin chains with baths coupled at the
ends of chain~\cite{PP-08,Prosen-08,Znidaric-15,SK-20}, where the
Liouvillian gap behaves as $\Delta_{\cal L} \sim L^{-3}$.
Indeed, we find that in the large-$L$ limit
\begin{equation}
  L^3 \Delta_{\cal L} \approx F(\bar{\mu},w) \,.
  \label{decall}
\end{equation}
In particular the function $F(0,w)$ is nonmonotonic with a minimum for
$w\approx 1.7265$ in correspondence of the maximum of the absolute
value of the fermionic current, see Fig.~\ref{Jwlt}.  These results
imply that the asymptotic approach to stationarity in Kitaev wires
with dissipative pumping/decay boundary dissipation is associated with
a large time regime scaling as $t\sim L^3$.

In Figs.~\ref{Gcwasy} and \ref{Gcwasy1b} we show some results for the
fermionic correlation in the case of boundary dissipations at both
ends and at only one end, respectively. They definitely support the
asymptotic dynamic scaling in the terms of the scaling time variable
\begin{equation}
  \Theta=t L^{-3}\,.
  \label{thetaldef}
  \end{equation}
This regime is again controlled by the interplay between the
Hamiltonian parameter $\bar{\mu}$ and the dissipative rate $w$.  The
results hint at the dynamic scaling behavior
\begin{eqnarray}
\widetilde{G}_{c/p}(x,y,\bar{\mu}_i,\bar{\mu},w,t,L) \approx
\widetilde{G}_{c/p}(X,Y,\kappa_i,\kappa,\Theta, w)\,.
\label{gcpscadiss4}
\end{eqnarray}
Note that within this large-time regime the system looses the memory
of the initial the critical condition of the system, approaching
a noncritical stationary state.

\section{Conclusions}
\label{conclu}

We have investigated how the presence of boundary dissipative
interactions (see Fig.~\ref{fig:sketchbou}) affects the quantum
critical dynamics of many-body systems at CQTs, i.e. when the
Hamiltonian parameters driving the unitary dynamics get tuned to their
critical values, leading to a vanishing gap and a diverging length
scale.

As a paradigmatic model, we consider the quantum fermionic
Kitaev wires, defined by the Hamiltonian (\ref{kitaev2}), and subject
to dissipative interactions at the boundaries, associated with
particle pumping and decay mechanisms. They are induced by couplings
with a Markovian bath, such that the evolution of the system density
matrix can be effectively described by a Lindblad master equation,
such as Eq.~(\ref{lindblaseq}).  The Kitaev wire with pumping/decay
dissipative interactions is particularly convenient for numerical
computations, indeed it allows us to perform numerical computations
for relatively large systems, and therefore to achieve accurate checks
of the dynamic scaling behaviors in the presence of dissipative
interactions with the environment, see also
Refs.~\onlinecite{NRV-19,RV-19,RV-20,DRV-20}.  In our study we address
the relevance of dissipative boundaries at CQTs, i.e.  whether they
maintain the system within a critical regime, or they make the system
depart from criticality. Moreover, we check if their effects can be
casted within a dynamic scaling framework as in the case of homogenous
dissipative mechanisms.

To address the quantum dynamic resulting from the competition of the
unitary Hamiltonian and boundary dissipative drivings, we consider
protocols (see Sec.~\ref{protocol}) based on an instantaneous
quenching of the Hamiltonian parameters and turning on of the
dissipative interactions, starting at $t=0$ from ground states of the
Hamiltonian with parameters close to their critical values.  Analogous
protocols were also considered to address the effects of homogenous
dissipative interactions involving the bulk of the
system~\cite{NRV-19,RV-19,RV-20,DRV-20}, as sketched in
Fig.~\ref{fig:sketchhom}, so that we can make an interesting
comparison of the effects of bulk and boundary dissipations described
within the analogous Lindblad framework.

On the one hand, in the case of bulk homogenous dissipation at quantum
transitions, the quantum dynamics of systems of size $L$ can be
described within dynamic FSS frameworks where the relevant scaling
variable associated with time is $\theta = t L^{-z}\sim t \Delta_L$
(where $\Delta_L\sim L^{-z}$ is the vanishing gap of the critical
Hamiltonian), and the global dissipative rate $u$ must be tuned to
zero as $u\sim \Delta_L\sim L^{-z}$ with $z=1$. The out-of-equilibrium
dissipative quantum dynamics shows essentially one dynamic regime,
from the beginning to the large-time asymptotic behavior.

On the other hand, quantum fermionic wires with boundary dissipations
show notable differences. In particular their quantum evolution during
the above mentioned protocol show two different dynamic regimes.
There is an early-time regime for times $t\sim L$, where the
competition between coherent and incoherent drivings develops a
dynamic FSS analogous to that applying to bulk dissipations, but the
boundary dissipative-rate parameter $w$ does not require to be tuned
to zero. Then there is a large-time regime for $t\sim L^3$ whose
dynamic scaling describes the late quantum evolution leading to the
$t\to\infty$ stationary states. The large time scales $t\sim L^3$ are
essentially related to the slowest decay of the Lindbladian gap
$\Delta_{\cal L}\sim L^{-3}$, which characterize several quantum spin
chains and fermionic wires with boundary
dissipations~\cite{PP-08,Prosen-08,Znidaric-15,SK-20}.

We present various numerical results for systems with decay and
pumping dissipative interactions with equal dissipation rate at their
ends, and also dissipative interactions localized to one end only. The
emerging scaling scenarios appear similar, thus we believe that their
validity extends to more general situations with localized dissipative
interactions. For example one may consider periodic wires close to
quantum transitions with one, or more then one, localized dissipative
interactions with external sources.

Further investigations are called for, to achieve a deeper
understanding of the effects of boundary dissipative interactions at
quantum transitions.  In this respect, a crucial role is played by the
exponent $\zeta$ entering the scaling law (\ref{gcpscadiss3}), and
controlling the scaling of the boundary dissipation parameters. Our
numerical results show that it is compatible with zero in fermionic
Kitaev wires with pumping and decay boundary dissipative mechanisms.
An interesting question is whether it assumes different values in
other one-dimensional models with boundary dissipations, which may be
also related to mechanisms that are not describable by Lindblad
equations, such as baths with an infinite set of harmonic
oscillators~\cite{CL-83,Leggett-etal-87}. Other interesting issues
concern higher-dimensional systems with dissipative interactions
around the boundaries. Moreover, one may also address the effects of
boundary dissipations at first-order quantum transitions, which are
characterised by an extreme sensitivity to the boundary
properties~\cite{CPV-14,PRV-18,PRV-20,DRV-20}.

\appendix

\section{Some details on the numerical computations}
\label{appnum}

To compute the time evolution of an observable $A(t)$
associated with an operator $\hat{A}$, 
\begin{equation}
  A(t) = {\rm Tr}[\rho(t) \,\hat{A}]\,, 
  \label{otdef}
  \end{equation}
  such as those defined in Sec.~\ref{protocol}, we solve corresponding
  coupled differential equations, formally obtained from the Lindblad
  master equation (\ref{lindblaseq}), as
\begin{eqnarray}
  {\partial \over \partial t}A(t) = {\rm Tr}[{\cal
    L}(\rho)\,\hat{A}]\,,\qquad
  A(0) = {\rm Tr}[\rho_i \,\hat{A}]\,.
\label{plieq}
\end{eqnarray}
To the purpose of computing the observables introduced in
Sec.~\ref{protocol}, we consider the quantities
\begin{eqnarray} 
  \mathscr{C}_{x,y}(t) = {\rm Tr}[\rho(t)\, \hat c_x^\dagger \hat
    c_y^{\phantom\dagger}] \,,\quad
\mathscr{P}_{x,y}(t) = {\rm
    Tr}[\rho(t)\, \hat c_x^\dagger \hat c_y^\dagger]
\,.\quad
\label{suppqua}
\end{eqnarray}
Then, straightforward computations allow us to derive the linear
equations
\begin{eqnarray}
&&  \frac{d}{dt}\,\mathscr{C}_{x,y} = i\,(\mathscr{C}_{x,y+1} -
  \mathscr{C}_{x-1,y} + \mathscr{C}_{x,y-1} - \mathscr{C}_{x+1,y})
  \qquad \label{eqscxy}\\
&&\quad -i \, (\mathscr{P}_{y,x-1}^\dagger
  - \mathscr{P}_{y,x+1}^\dagger - \mathscr{P}_{x,y-1} + \mathscr{P}_{x,y+1})
  \nonumber \\
  & &  \quad
  - \frac{w}{2} \,( \delta _{L,y}\,
  \mathscr{C}_{x,L} + \delta _{L,x}\, \mathscr{C}_{L,y} 
+\delta_{1,y}\, \mathscr{C}_{x,1} + \delta _{1,x}\,
  \mathscr{C}_{1,y}) \nonumber\\
&&\quad + w \,\delta_{L,y}\,\delta_{L,x} \,,\nonumber\\
&&\frac{d}{dt}\,\mathscr{P}_{x,y} = -i\,(\mathscr{P}_{x,y+1} +
    \mathscr{P}_{x+1,y}+ \mathscr{P}_{x,y-1} + \mathscr{P}_{x-1,y})
    \nonumber\\
 && \quad -
i\,( \mathscr{C}_{x,y-1} -
\mathscr{C}_{y,x-1} - \mathscr{C}_{x,y+1} 
+ \mathscr{C}_{y,x+1}) \nonumber \\
 && \quad -
i\,(\delta _{x-1,y} - \delta_{x+1,y})
- 2\,i\,\mu \,\mathscr{P}_{x,y} \nonumber \\
&&\quad - \frac{w}{2} \,(\delta_{1,y}\,\mathscr{P}_{x,1} +
\delta_{1,x}\,\mathscr{P}_{1,y} + \delta _{L,y}\,\mathscr{P}_{x,L}
+ \delta_{L,x}\,\mathscr{P}_{L,y}) \,.  
\nonumber
\end{eqnarray}
These coupled equations must be solved using the initial conditions
$\mathscr{C}_{x,y}(0) = {\rm Tr}[\rho_i\, \hat c_x^\dagger \hat
  c_y^{\phantom\dagger}]$ and $\mathscr{P}_{x,y}(0) = {\rm
  Tr}[\rho_i\, \hat c_x^\dagger \hat c_y^\dagger]$, where $\rho_i$ is
the initial pure-state density matrix corresponding to the ground
state of the Hamiltonian for $\bar{\mu}_i$.  They can be computed
using standard diagonalization techniques, see, e.g.,
Ref.~\onlinecite{Blaizot-book}.  Then differential equations are
solved using the four-order Runge-Kutta method.  Finally, the
observables defined in Sec.~\ref{protocol} are easily obtained by
$D(x,t) = \mathscr{C}_{x,x}(t)$, $J(x,t) = -2\,{\rm Im}
\,\mathscr{C}_{x+1,x}(t)$, $G_c(x,y,t) = 2\,{\rm
  Re}\,\mathscr{C}_{x,y}(t)$, and $G_p(x,y,t) = 2\,{\rm
  Re}\,\mathscr{P}_{x,y}(t)$.

We finally describe how we obtained the asymptotic stationary
behaviors reported in Eqs.~(\ref{dasy}), (\ref{jasy}), and
(\ref{gcpasy}).  First, we solved Eqs.~(\ref{eqscxy}) in the
stationary limit for systems of finite size $L$ using exact
diagonalization, by assuming that the time derivatives in the
l.h.s. vanish.  The results turn out to rapidly converge to their
large-$L$ limit, as for example shown by the data reported in
Fig.~\ref{Jwlt}.  This allows us to obtain a robust guess of their
large-$L$ limits, such as those reported in Eqs.~(\ref{dasy}),
(\ref{jasy}), and (\ref{gcpasy}). Then we verified that they exactly
solve the coupled equations in the stationary and large-$L$ limits.
Of course, these results are consistent with the asymptotic large-time
convergence of the observables in the time evolution arising from the
dynamic protocol.

\end{document}